\documentclass[aps,prb,twocolumn,superscriptaddress,showpacs]{revtex4-1}
\usepackage{graphicx}
\usepackage{amsmath,amssymb,bm}
\usepackage{times}
\usepackage[colorlinks,citecolor=blue,linkcolor=red]{hyperref}
\usepackage{color}

\begin{document}

\title{Asymmetric Andreev Reflection Induced Electrical and Thermal Hall-like Effects in Metal/Anisotropic Superconductor Junctions}

\author{Jie Ren}\email{renjie@lanl.gov}
\affiliation{Theoretical Division, Los Alamos National Laboratory, Los Alamos, New Mexico 87545, USA}

\author{Jian-Xin Zhu}\email{jxzhu@lanl.gov}
\affiliation{Theoretical Division, Los Alamos National Laboratory, Los Alamos, New Mexico 87545, USA}
\affiliation{Center for Integrated Nanotechnologies, Los Alamos National Laboratory, Los Alamos, New Mexico 87545, USA}


\begin{abstract}
By investigating the nonequilibrium transport across a metal/superconductor junction in both nonrelativistic and relativistic cases, we reveal that the asymmetric Andreev reflection with anisotropic superconductors is able to induce electric and thermal Hall-like effects, in the absence of a magnetic field. That is, a longitudinal electric voltage or temperature bias can induce transverse electric or thermal currents merely through the asymmetric Andreev reflection, respectively. 
In particular, a transverse thermoelectric effect, i.e., the Ettingshausen-like effect, is identified, although the conjugate Nernst effect is absent. The direction change of these electric and thermal Hall-like currents is also discussed. 
The Hall-like effects uncovered here do not require the conventional time-reversal symmetry breaking, but rather originate  from the mirror symmetry breaking with respect to the interface normal due to the anisotropic paring symmetry of the superconductor.
\end{abstract}


\maketitle

\section{Introduction}

It was first noted by Alexander~F.~Andreev 50 years ago that the interface between a metal and a superconductor can retro-reflect an incident electron as a positive charged hole, while the missed charges enter the superconductor as a Cooper pair, thereafter called Andreev reflection (AR).~\cite{Andreev} Since its disclosure, the AR has never stopped surprising us with new phenomena, such as a zero-bias conductance peak~\cite{ZBCP, BTK2} in $d$-wave superconductors, non-local crossed AR,~\cite{CAR1, CAR2} specular AR in Dirac materials,~\cite{Beenakker, DYXing}  Majorana fermions formed by exotic Andreev bound states in topological superconductors~\cite{Fu}. In this work, we report a new phenomenon related to AR, that is, the asymmetric Andreev reflection (AAR) in anisotropic superconductors is able to induce electric and thermal Hall-like effects, in the absence of a magnetic field.~\cite{like-mean}

Conventionally, various classical and quantum Hall effects arise in the presence of magnetic orders, like magnetization and an external magnetic field. In other words, Hall effects usually require broken time-reversal symmetry. One exception is the various spin-related Hall effects with spin-orbital-like interactions in the absence of magnetic fields,~\cite{Hasan2010RMP, Qi2011RMP} which can occur with the time-reversal symmetry preserved because the spin current is even under time-reversal operation.
A recent study also showed that with the application of space-time dependent potentials on a Dirac material, a Hall-like charge current emerges.~\cite{savelev2012PRL} In addition, the Hall-like thermoelectric transport is also proposed as a new and promising direction for microscale and cryogenic Peltier cooling,~\cite{HallTE} wherein a longitudinal electric current generates a transverse heat current without a magnetic field.

The AAR that we uncover here provides a novel mechanism for Hall-like transverse charge and heat  transports, in spite of the fact that 
the transverse momentum is conserved across the junction. It has direct implications for transverse microscale and cryogenic Peltier cooling. The AAR-induced Hall-like effects do not require broken time-reversal symmetry. They result from the breaking mirror symmetry with respect to the interface normal due to the anisotropic pairing of the superconductor.

\section{Model and Results}

\begin{figure}
\scalebox{0.31}[0.31]{\includegraphics{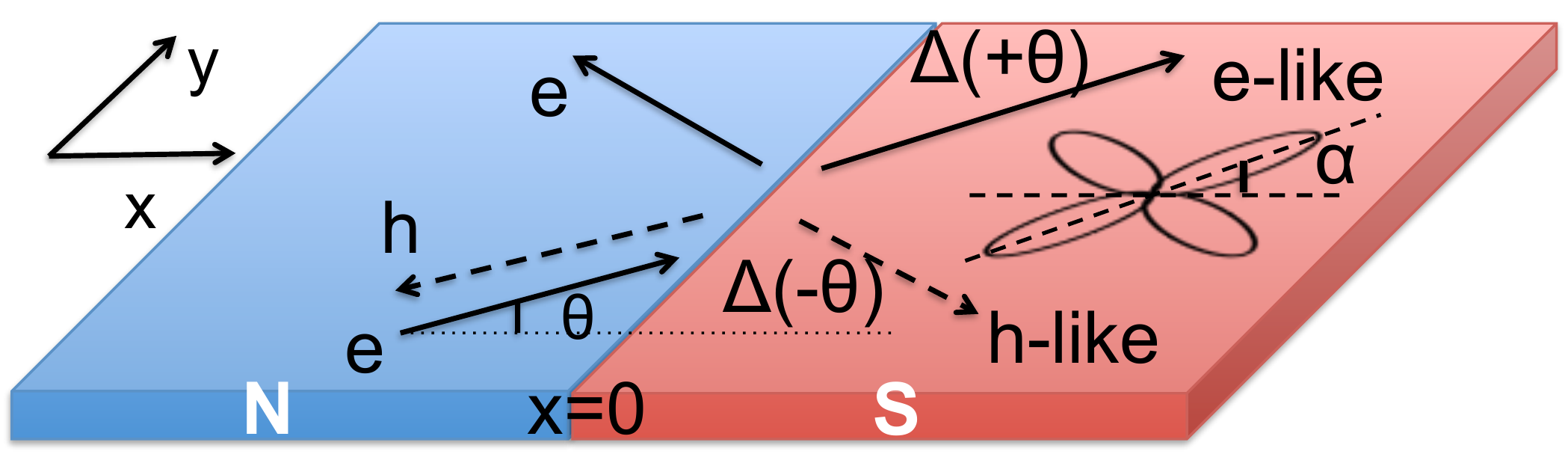}}
\vspace{-0.3cm}
\caption{{\bf Schematic of the Andreev reflection, normal reflection, and transmissions at the N/S interface.} In the nonrelativistic case, the N side is just the topologically trivial normal metal. In the relativistic case, the N side can be graphene, the metallic surface of topological insulators, or other Dirac-like materials. The S side is the corresponding superconductor with $d$-wave symmetry.}
\label{fig:scheme}
\end{figure}

Let us consider a two-dimensional metal/superconductor (N/S) junction, with the $y$-direction translational invariant interface located at $x=0$, which is assumed to be the only boundary of the problem. Therefore, we analyze the bulk transport, where the elementary (quasi-)particles with energy $E$ are described by the Bogoliubov-de Gennes equation~\cite{BdGbook} 
\begin{equation}
\hat{H}\binom{f}{g}=E\binom{f}{g}
\end{equation}
 with the total Hamiltonian
\begin{eqnarray}
\hat{H}\!=\!\left(
        \begin{array}{cc}
       \!\!   \hat{H}_0(\vec{k})+V(x)-E_{F} \!\!&\!\! \hat\Delta\Theta(x) \!\! \\
      \!\!    \hat\Delta^{\dag}\Theta(x) \!\!&\!\! E_{F}-\hat{H}^*_0(-\vec{k})-V(x) \!\! \\
        \end{array}
      \right).
    \label{eq:BdG}
\end{eqnarray}
$f$ and $g$ denote the electron-like and hole-like components of the wave function, respectively.  $\hat{H}_0(\vec{k})$ is the Hamiltonian for electrons on the N side while $\hat{H}^*_0(-\vec{k})$ is the time-reversal counterpart for holes. $V(x)$ denotes the barrier potential at the interface. The Fermi energies $E_F$ are assumed to be equal on both the N and the S sides. Without loss of generality, the step-like superconducting pair potential $\hat\Delta\Theta(x)$ is adapted so that the region $x<0$ denotes the N side while the region  $x>0$ represents the S side.~\cite{Tinkhambook} Note that in the weak coupling limit, the superconducting coherence length is much larger than the Fermi wavelength,~\cite{BdGbook} $E_F\gg |\Delta|$. 

We consider the S side to have a $d$-wave pairing symmetry, representative of anisotropic superconductors,  and assume electrons  to be injected from the N side with incoming angle $\theta$ and energy $E>0$ measured from $E_F$. As depicted in Fig.~\ref{fig:scheme}, the reflection at the interface is either a normal specular-reflection with amplitude $b(E,\theta)$ or an Andreev retro-reflection, converting electrons to holes, with amplitude $a(E,\theta)$. The transmitted electron- and hole-like quasiparticles experience two different values of pair potential, $\Delta(\theta)=\Delta_0\cos(2\alpha-2\theta)$ and $\Delta(-\theta)=\Delta_0\cos(2\alpha+2\theta)$, with amplitudes $c(E,\theta)$ and $d(E,\theta)$, respectively, where $\Delta_0$ is the maximum of the angle-dependent superconducting gap and $\alpha$ denotes the mis-oriented angle between the crystal axis of the $d_{x^2-y^2}$-wave superconductor and the normal direction of the interface.

For both reflection and transmission processes, the transverse ($y$-direction) momentum parallel to the interface is conserved across the interface.~\cite{Andreev} Therefore, for incident electrons  with vanishing $y$-momentum on average over $\theta$, one would never expect that a nonvanishing transverse transport emerges. 
Therefore, the conventional wisdom (as, indeed, is the case in the literature and textbooks) has never considered Hall transports in the absence of a magnetic field since the disclosure of AR a half-century ago.~\cite{Andreev} This situation remains even after the discovery of anisotropic superconductors. 
However, as we show below, $|\Delta(\theta)|\neq|\Delta(-\theta)|$ as a peculiar feature of anisotropic superconductors gives rise to the AAR amplitude $(|a(\theta)|^2\neq|a(-\theta)|^2)$, which subsequently induces the anomalous electric and thermal Hall-like effects in the transverse direction. 
We  exemplify these AAR-induced Hall-like effects in both a nonrelativistic and a relativistic  case, adapting the Andreev condition $E_F\gg(|\Delta|,E)$ without loss of generality.

\subsection{Nonrelativistic asymmetric-Andreev-reflection.} 
For the nonrelativistic case where the N side is a normal metal with the S side a superconductor, we have $(f,g)$ acting on the basis $(\psi_e,\psi_h)$ and $\hat{H}_0(\vec{k})={\hbar\vec{k}^2}/{(2m)}$. The interface barrier is modeled as a delta-function $V(x)=U_0\delta(x)$ and $\hat\Delta=\Delta(\theta)=\Delta_0\cos(2\alpha-2\theta)$ denotes the angle-dependent superconducting gap.

Since the electrons and holes are both the energy carriers but with opposite charge, we define the energy probability density $\rho_Q=E(|f|^2+|g|^2)$ and the charge probability density $\rho_e=e(|f|^2-|g|^2)$. By substituting these definitions into the Bogoliubov-de Gennes equation, we have  two conservation laws:~\cite{BTK00, BTK0, BTK1}
\begin{subequations}\label{eq:conservation}
\begin{align}
&\frac{\partial\rho_Q}{\partial t}+ \vec \nabla \cdot \vec J_Q=0;  \\
&\frac{\partial\rho_e}{\partial t}+ \vec \nabla \cdot \vec J_e=\frac{4e}{\hbar}\text{Im}[f^{\dag}\hat\Delta g], 
\end{align}
\end{subequations}
which  determine the energy current density $\vec{J}_Q=\frac{E\hbar}{m}\text{Im}[f^{\dag}\vec\nabla{f}-g^{\dag}\vec\nabla{g}]$ and the electric current density $\vec{J}_e=\frac{e\hbar}{m}\text{Im}[f^{\dag}\vec\nabla{f}+g^{\dag}\vec\nabla{g}]$. Since currents are conserved across the junction interface, we can simply consider the wave functions only on the N side:~\cite{BTK1}
\begin{equation}
\binom{f}{g}=\left[\binom{1}{0}e^{ik_xx}+a\binom{0}{1}e^{ik_xx}+b\binom{1}{0}e^{-ik_xx}\right]e^{ik_yy},
\label{eq:wave}
\end{equation}
where $k_x=k_F\cos\theta$ and $k_y=k_F\sin\theta$ with $k_F=\sqrt{2mE_F}/\hbar$. The first term on the right-hand side denotes the incoming wave function, the second term denotes the AR converting electrons to holes with amplitude $a(E, \theta)$ and the last term is the normal reflection with amplitude $b(E,\theta)$. 
Substituting it into the transverse current formulas $J_Q^y=\frac{E\hbar}{m}\text{Im}(f^{\dag}\partial_yf-g^{\dag}\partial_yg)$ and $J_e^y=\frac{e\hbar}{m}\text{Im}(f^{\dag}\partial_yf+g^{\dag}\partial_yg)$, we then find the angle-resolved $y$-direction energy and electric current densities:
\begin{eqnarray}
J_Q^y&=&2Ev_F\sin\theta\left(1-|a|^2+|b|^2+2\text{Re}[b e^{-2ik_xx}]\right); \\
J_e^y&=&2ev_F\sin\theta\left(1+|a|^2+|b|^2+2\text{Re}[b e^{-2ik_xx}]\right),
\end{eqnarray} 
where $v_F=\hbar k_F/m$ and the prefactor 2 is restored for the spin degrees of freedom. The opposite signs before $|a|^2$ in $J^y_Q$ and $J^y_e$ indicate that the AR blocks (retro-reflects) the energy transport but transmits the charge by generating a Copper pair on the S side. The positive $|b|^2$ implies that the specular normal reflection transmits the $y$-component of both energy and electric currents. The last term $2\text{Re}[b e^{-2ik_xx}]$ results from the interference between the incident wave and the normal reflected one, which produces transverse currents that are oscillatory-dependent on the $x$-position.

However, as we will see, after considering all the possibilities of the incoming angle $\theta$, all the terms containing $b(E, \theta)$ will vanish and only the term containing AR $a(E, \theta)$ will survive. The normal reflection can not provide the transverse transport on average, but the AAR can. In fact, if we follow Ref.~\cite{BTK2} to solve $a$ and $b$, we get 
\begin{eqnarray}
a(E, \theta)&=& \frac{\cos^2\theta\Gamma_+e^{-i\phi_+}}{\cos^2\theta+Z^2(1-\Gamma_+\Gamma_-e^{-i(\phi_+-\phi_-)})},   \\
b(E, \theta)&=& \frac{-Z(Z+i\cos\theta)(1-\Gamma_+\Gamma_-e^{-i(\phi_+-\phi_-)})}{\cos^2\theta+Z^2(1-\Gamma_+\Gamma_-e^{-i(\phi_+-\phi_-)})},
\end{eqnarray}
where $Z={U_0}/{(\hbar{v_F})}$ is the dimensionless barrier strength,~\cite{imperfectness} $e^{i\phi_{\pm}}=\Delta(\pm\theta)/|\Delta(\pm\theta)|$ and $\Gamma_{\pm}=v_{\pm}/u_{\pm}$,~with $v_{\pm}=\sqrt{\frac{1}{2}(1-{\sqrt{1-|\Delta(\pm\theta)|^2/E^2}})},~u_{\pm}=\sqrt{\frac{1}{2}(1+{\sqrt{1-|\Delta(\pm\theta)|^2/E^2}})}$.
From the expressions, we know that $b$ is an even function of $\theta$, so that after integration over the incident angle $\theta$, the contributions of $|b|^2\sin\theta$ and $\text{Re}[be^{-2ik_xx}]\sin\theta$ to both current densities will vanish. However, $a$ is not an even function of $\theta$ because $\Gamma_+\rightarrow\Gamma_-$ under $\theta\rightarrow-\theta$. Therefore,  AAR plays a crucial role in the Hall-like effect in anisotropic superconductors.

\subsection{Relativistic asymmetric-Andreev-reflection.} 
For the relativistic case, the N side is a Dirac material such as graphene or the metallic surface of a topological insulator, and the S side is the corresponding superconductor. Taking the latter (the metallic surface of a topological insulator) for example, we have $(f, g)$ acting on the Nambu basis $(\psi_{e\uparrow},\psi_{e\downarrow},\psi_{h\uparrow},\psi_{h\downarrow})$ and $\hat{H}_0(\vec{k})=\hbar v_{F}(k_x\hat\sigma_y-k_y\hat\sigma_x)$, where $v_F$ is the Fermi velocity and $\hat\sigma_{x(y)}$ denote Pauli matrices.~\cite{Hasan2010RMP, Qi2011RMP} Using graphene will make a small difference in the mathematical representation, but the physics and final results will not change. The interface barrier is described by $V(x)=\frac{U_0}{d}\Theta(x)\Theta(x-d)$ with $d$ being the barrier thickness. By taking the limit~\cite{Bhattacharjee2006PRL} $d\rightarrow0$, we obtain a dimensionless barrier strength $Z=U_0/(\hbar{v_F})$. The angle-dependent superconducting pair potential now becomes $\hat\Delta=i\hat\sigma_y\Delta(\theta)\Theta(x)$. 

In this relativistic case, $f=\binom{\psi_{e\uparrow}}{\psi_{e\downarrow}}$ and $g=\binom{\psi_{h\uparrow}}{\psi_{h\downarrow}}$, and we still have the same definitions of the energy and charge probability density  as well as the same conservation laws~Eqs.~(\ref{eq:conservation}). Because of the distinct Dirac nature of the $\hat H_0(\vec k)$, now we have different current density expressions,~\cite{Ren2013PRB} whose transverse components read $J_Q^y=-Ev_F(f^{\dag}\sigma_xf+g^{\dag}\sigma_xg)$ and $J_e^y=-ev_F(f^{\dag}\sigma_xf-g^{\dag}\sigma_xg)$. Considering the wave function on the N side:
\begin{eqnarray}
\binom{f}{g}=\left(\begin{array}{c}e^{ik_xx}+be^{-ik_xx}\\ie^{i\theta}e^{ik_xx}-ibe^{-i\theta}e^{-ik_xx}\\aie^{ik_xx}\\-ae^{-i\theta}e^{ik_xx}\end{array}\right)e^{ik_yy},
\end{eqnarray}
where $k_x=k_F\cos\theta$ and $k_y=k_F\sin\theta$ with $k_F=E_F/(\hbar v_F)$,
we then obtain the angle-resolved transverse energy and electric current densities:
\begin{align}
J_Q^y \!&=\! 2Ev_F \! \left(\sin\theta(1 \!-\! |a|^2 \!+\! |b|^2) \!+\! 2\text{Im}[b^* e^{i(\theta+2k_xx)}]\right);  \\
J_e^y \!&=\! 2ev_F \! \left(\sin\theta(1 \!+\! |a|^2 \!+\! |b|^2) \!+\! 2\text{Im}[b^* e^{i(\theta+2k_xx)}]\right).
\end{align} 
The interference term $\text{Im}[b^*e^{i(\theta+2k_xx)}]$ is different from that in the nonrelativistic case. Despite this difference, we will see that all terms containing $b$ will vanish after considering all possible $\theta$ and only AAR has the possibility of giving rise to transverse transport,  just the same as in the nonrelativistic case.
Following Ref.~\onlinecite{Ren2013PRB} to solve $a(E,\theta)$ and $b(E,\theta)$, we arrive at
\begin{align}
a&=\frac{-\cos^2\theta \Gamma_+ e^{i(\theta-\phi_+)}}{\cos^2\theta+\sin^2Z\sin^2\theta(1-\Gamma_+\Gamma_-e^{i(\phi_--\phi_+)})},    \\
b&=\frac{\sin Z\sin\theta(\cos Z\cos\theta-i\sin Z)(1-\Gamma_+\Gamma_-e^{i(\phi_--\phi_+)})}{-e^{-i\theta}[\cos^2\theta+\sin^2Z\sin^2\theta(1-\Gamma_+\Gamma_-e^{i(\phi_--\phi_+)})]},
\end{align}
where the barrier strength $Z$ confined by trigonometric functions is a manifestation of the relativistic Klein tunneling.~\cite{Ren2013PRB,Klein1,Klein2} Other parameters have the same meanings as in the nonrelativistic case.
Since $|b(E,\theta)|^2$ is symmetric (even function) with respect to $\theta$, the contribution of $|b|^2\sin\theta$ will be 0 after integration over $\theta$. Moreover, $\text{Im}[b^*e^{i(\theta+2k_xx)}]$ is antisymmetric (odd function) with respect to $\theta$ so that after integration its contribution is 0 as well. However, $|a(E,-\theta)|\sim|\Gamma_-|$ is not equal to $|a(E,\theta)|\sim|\Gamma_+|$ generally so that $|a|^2\sin\theta$ survives after angle averaging. Therefore, only the AAR is able to contribute to the emergence of transverse Hall-like currents. 

\subsection{Electric and Thermal Hall-like effects.} 
\begin{figure}
\scalebox{0.36}[0.36]{\includegraphics{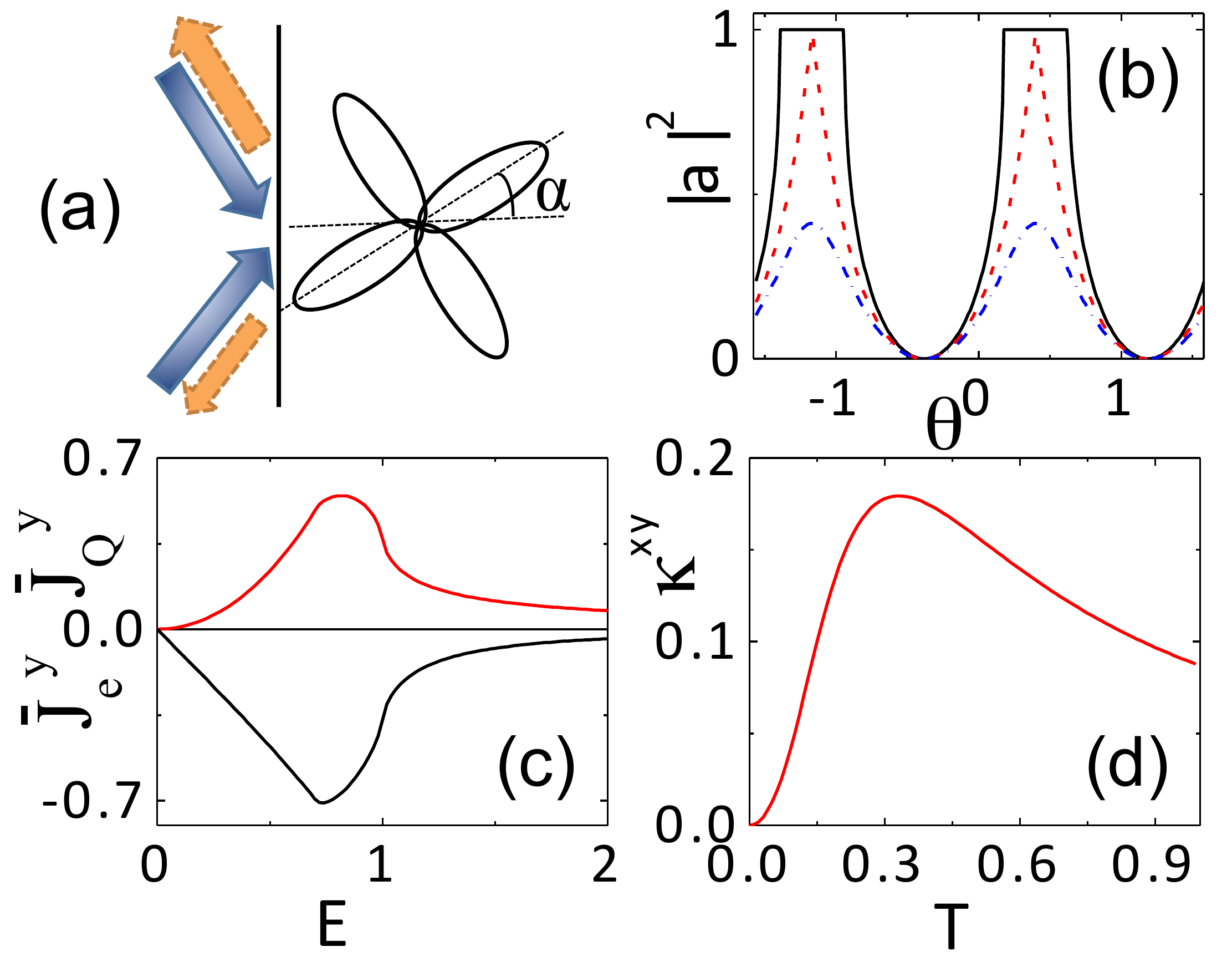}}
\vspace{-.5cm}
\caption{{\bf AAR and induced electric and thermal Hall-like effects.} (a) Schematic of the AAR.  ARs with opposite incident angles have asymmetric amplitudes. (b) The angle-resolved AR amplitudes clearly show the asymmetry with respect to the normal incidence $\theta=0$, for $E=0.9$ (solid line), $1.0$ (dashed line), and $1.1$ (dash-dotted line). (c) Transverse thermal and electric currents as a function of energy. (d) Thermal Hall conductance as a function of temperature. Other parameters are $\alpha=\pi/8,\Delta_0=1,e=1,v_F=1$, and $k_B=1$. We set $Z=0$, so that both nonrelativistic and relativistic cases have $|a|^2=|\Gamma_+|^2$ and behave the same. The  constant $N_{E_F}$ is normalized.}
\label{fig:nhall}
\end{figure}

From the above discussion, we know that normal reflection does not play any role in the possible Hall-like effects. After considering all possible incident angles, we then have the expressions of the total transverse Hall-like current densities for both nonrelativistic and relativistic cases: 
\begin{eqnarray}
\bar J_e^y(E) \!&=&\! \!\! \int^{\pi/2}_{-\pi/2} \!\!\!\!\!\! d\theta J_e^y(E, \theta) \!= \! 2ev_F \!\! \int^{\pi/2}_{-\pi/2} \!\!\!\!\!\! d\theta \sin\theta |a|^2, \\
\bar J_Q^y(E) \!&=&\! \!\! \int^{\pi/2}_{-\pi/2} \!\!\!\!\!\!  d\theta J_Q^y(E, \theta) \! =\! - 2Ev_F \!\! \int^{\pi/2}_{-\pi/2} \!\!\!\!\!\!\ d\theta \sin\theta |a|^2.
\end{eqnarray} 
To observe these transverse Hall-like transport properties, the crystalline angle of the anisotropic superconductor should be oriented to avoid the integral multiple of $\alpha=\pi/4$ so that the angle-dependent superconducting gap $\Delta_0\cos(2\alpha-2\theta)$ as well as the AR $|a|^2$ is asymmetric with respect to the normal incident angle $\theta=0$ [see Fig. ~\ref{fig:nhall}(a) and (b)]. This actually indicates that the mirror symmetry breaking with respect to the interface normal is required for manifestation of the AAR-induced Hall-like currents. 
The induced Hall-like effects are illustrated in Fig.~\ref{fig:nhall}(c).
With increasing the energy, the electric and thermal Hall-like effects decrease to 0 because the AR tends to disappear when the quasiparticle energy is much beyond the superconducting gap. The opposite sign between $\bar{J}_Q^y(E)$ and $\bar{J}_e^y(E)$ is a consequence of the electron-hole converting nature of AR, wherein the reflected holes carry energy and charge currents in opposite directions. 
As we show below, these two quantities are also the differential conductances at voltage bias $eV=E$ at low temperatures.  

The conventional BTK formula~\cite{BTK1, BTK2, Beenakker} concerning the longitudinal electric current, reads $I^x_e=N_{E_F}\int{dE}\bar{J}^x_e(f_L-f_R)$, with $\bar{J}^x_e=2ev_F\int{d\theta}\cos\theta(1+|a|^2-|b|^2)$ being the total longitudinal electric current density and $N_{E_F}$ the density of states near the Fermi level, which can be pulled out of the integral in the wide band approximation. Clearly, all incoming currents from the N side have the Fermi distribution function $f_L=[e^{(E-eV)/(k_BT_L)}+1]^{-1}$, while those coming in from the S side are weighted by $f_R=[e^{E/(k_BT_R)}+1]^{-1}$. Similarly, the transverse current contributed by the left-to-right current will be (partially) canceled by the opposite one contributed by the right-to-left current. Therefore, for the Hall-like electric current, we have a similar expression:
\begin{eqnarray}
I^y_e&=&N_{E_F}\int dE \bar{J}^y_e(E)(f_L-f_R)   \nonumber\\
&=&2ev_F N_{E_F}\!\! \int \!\!  dE\!\! \int\!\! d\theta  \sin\theta |a|^2(f_L-f_R).
\label{eq:electronHall}
\end{eqnarray}
The longitudinal heat current has a similar BTK-type expression:~\cite{Ren2013PRB, heatBTK1, heatBTK2}  $I^x_Q=N_{E_F}\int{dE}\bar{J}^x_Q(f_L-f_R)$ with $\bar{J}^x_Q=2Ev_F\int{d\theta}\cos\theta(1-|a|^2-|b|^2)$ being the total longitudinal energy current density. Accordingly, 
the transverse Hall-like heat current reads similarly:
\begin{eqnarray}
I^y_Q&=&N_{E_F}\int dE \bar{J}^y_Q(E)(f_L-f_R)   \nonumber\\
&=&-2v_F N_{E_F} \!\! \int \!\!  dE\!\! \int\!\! d\theta  E\sin\theta |a|^2(f_L-f_R).
\label{eq:heatHall}
\end{eqnarray}

Equations~(\ref{eq:electronHall}) and (\ref{eq:heatHall}) indicate that the electric and thermal Hall-like effects are observable under the nonequilibrium condition $f_L\neq{f}_R$ with either nonzero longitudinal voltage bias $eV$ or thermal bias $T_L\neq{T}_R$. Let us first examine the effect of voltage bias without thermal bias. In this case, at low temperatures, we get the differential Hall conductances $G^{xy}_e:={\partial{I}^y_e}/{\partial{eV}}=N_{E_F}\bar{J}^y_e(eV)$ and $G^{xy}_Q:={\partial{I}^y_Q}/{\partial{eV}}=N_{E_F}\bar{J}^y_Q(eV)$, which are shown in Fig.~\ref{fig:nhall}(c). Strictly speaking, the transverse heat current $I^y_Q$ induced by the longitudinal voltage bias under a magnetic field is called the Ettingshausen effect. Here, it is evident that even in the absence of a magnetic field, the AAR is able to induce the Ettingshausen-like effect.

We then turn to examine the effect of thermal bias without voltage bias.
Because of the AAR-induced Ettingshausen-like effect discussed above, one may expect that the AAR can also induce  the reversed process, Nernst-like effect, which is the transverse electric current $I^y_e$ induced by the longitudinal thermal bias.  However, due to the electron-hole symmetry of the AR, i.e., $a(E)=a(-E)$, it is straightforward to prove that Eq.~(\ref{eq:electronHall}) is always 0 with only thermal bias $T_L\neq{T}_R$, because $|a|^2(f_L-f_R)$ is an odd function of $E$ so that its integral over $E$ vanishes. Therefore, the AAR-induced Nernst-like effect is absent.~\cite{absent}
Nevertheless, the thermal Hall-like effect still survives, that is,  the longitudinal thermal bias is able to induce the transverse heat current $I^y_Q$. The corresponding thermal Hall conductance $\kappa^{xy}:=\partial{I}^y_Q/\partial\delta{T}=N_{E_F}\int{dE}\frac{E\bar{J}^y_Q}{4k_BT^2\cosh^2[E/(2k_BT)]}$ is displayed in Fig.~\ref{fig:nhall}(d). We here ignore the temperature dependence of $\Delta_0$: the higher the temperature the smaller $\Delta_0$. If considering such temperature dependence, the decrease in $\kappa^{xy}$ at high temperatures will be more pronounced.

\begin{figure}
\scalebox{0.42}[0.42]{\includegraphics{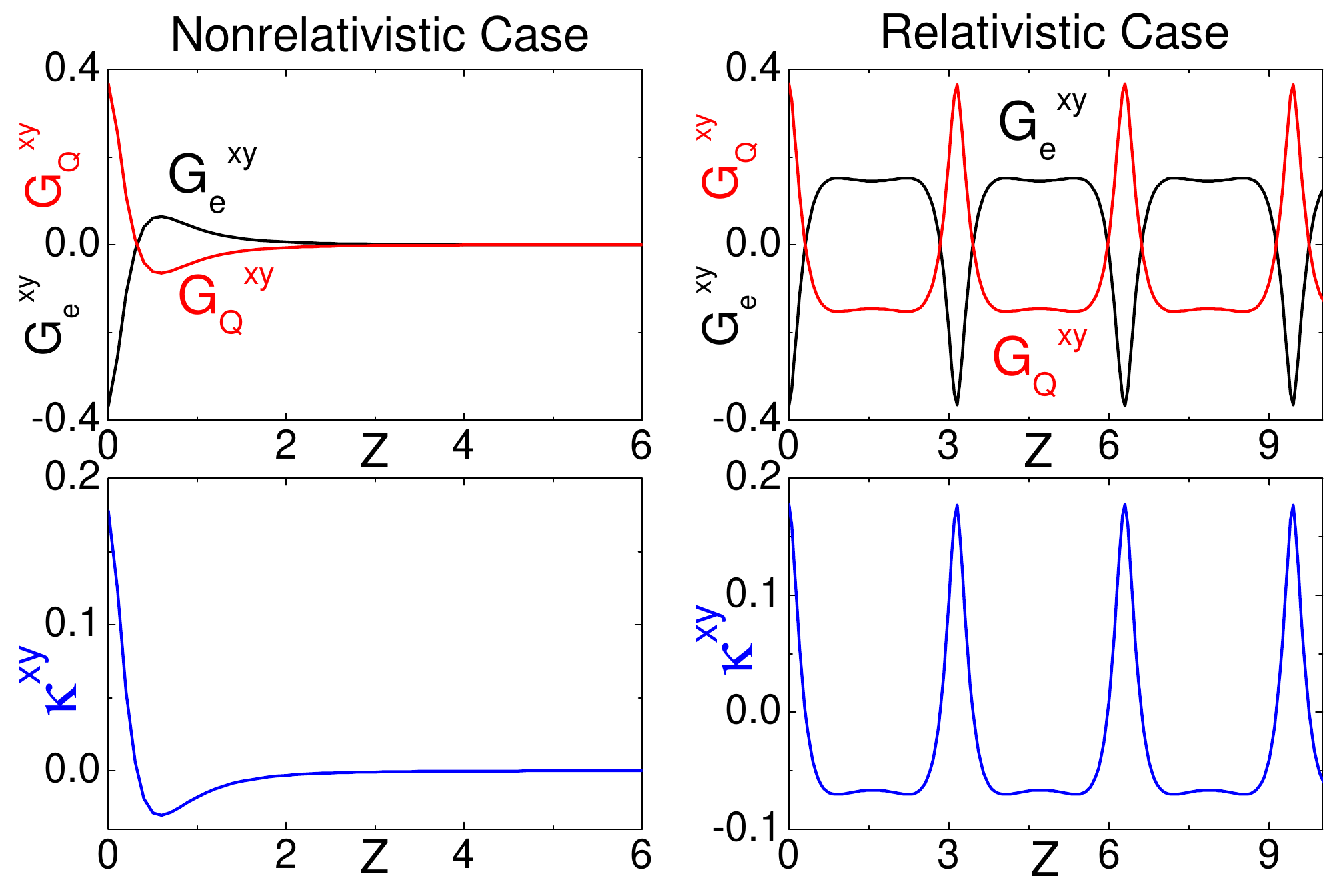}}
\vspace{-.5cm}
\caption{{\bf Barrier effect on AAR-induced Hall-like effects for both nonrelativistic (left) and relativistic (right) cases.} When calculating $G^{xy}_e:={\partial{I}^y_e}/{\partial{eV}}$~and~$G^{xy}_Q:={\partial{I}^y_Q}/{\partial{eV}}$, the bias $eV=1$ is fixed. When calculating $\kappa^{xy}:=\partial{I}^y_Q/\partial\delta{T}$, $T=0.3$ is fixed.  Other parameters are $\alpha=\pi/8,\Delta_0=1,e=1,v_F=1$, and $k_B=1$, with $N_{E_F}$ normalized.}
\label{fig:Z}
\end{figure}

Finally, we find that tuning the interface barrier is able to reverse the direction of Hall-like effects because the barrier influences the weight of AR amplitude $|a(\theta)|^2$ at positive and negative angles so that the angle integral of $\sin\theta|a(\theta)|^2$ may change the sign. For the nonrelativistic case, when increasing the barrier strength $Z$, the Hall-like effects first reverse their directions and then vanish to 0 because at large barrier the AR diminishes [left panel in Fig. \ref{fig:Z}]. However,  the direction and magnitude of Hall-like effects in the relativistic case show periodic oscillating behaviors via increasing $Z$ [right panel in Fig.~\ref{fig:Z}]. The oscillating behavior of the Hall-like effects as a function of the barrier $Z$ is a manifestation of Klein tunneling in relativistic transport.~\cite{Ren2013PRB} In addition, we note that the direction and magnitude of these electric and thermal Hall-like effects can also be tuned by changing either the crystal axis angle $\alpha$ or equivalently the range of incident angles.

\section{Conclusions and Discussions}

In summary, we have revealed that the AAR in anisotropic superconductors is able to induce electric and thermal Hall-like effects, in the absence of a magnetic field. A longitudinal electric voltage or temperature bias can induce transverse electric or thermal currents merely through the AAR, respectively. 
In particular, a transverse thermoelectric effect, i.e., the Ettingshausen-like effect, has been identified, which has direct implications in transverse cryogenic Peltier cooling.~\cite{HallTE} 
The direction change of these electric and thermal Hall-like currents has also been discussed.  
The Hall-like effects discussed here do not require the conventional time-reversal symmetry breaking, but result from the mirror symmetry breaking with respect to the interface normal due to the anisotropic paring symmetry of the superconductor. The spin-orbital interaction in the relativistic case is not crucial for the Hall-like effects, since the nonrelativistic case without spin-orbital interaction behaves similarly, except for the relativistic Klein tunneling effect. 

The AAR-induced Hall effects are observable under the condition that the crystal axis of the anisotropic superconductor is mirror-symmetry-broken, once the AR is significant. In fact, the $d$-wave AR has been widely observed in various experiments at metal/high $T_c$ superconductor junctions,~\cite{exp} even with imperfect, rough interfaces.
Moreover, for the relativistic case (such as graphene or the metallic surface of topological insulators), the imperfect interface will generally become transparent without severely affecting transport thanks to the Klein tunneling, see Fig.~\ref{fig:Z}. 
We thus believe AAR-induced Hall-like effects can be readily observed, once researchers turn their attention to the $y$-direction transport induced by AR.
Although we focus only on the N/S junction here, it would be interesting to explore the similar Hall-like Josephson supercurrent at S/N/S junctions with unconventional superconducting pairing symmetry.~\cite{JXZhu:1996,HXTang:1996} In fact, a recent experiment~\cite{nano_exp} measured the quantum Hall effect in nano-sized graphene with superconducting electrodes under a strong magnetic field.
In such experiments without an applied magnetic field, when one of the S parts is replaced with an anisotropic superconductor, we expect to observe an AAR-induced Hall-like supercurrent, if the Hall bar measurement setup is added. 

We caution that the present work is limited to the ballistic transport regime, since we have applied the scattering wave approach in the noninteracting single particle picture. As such, the studies are applicable to a clean sample, or a nano-sized junction (see Ref.~\onlinecite{nano_exp}) with weak impurities, where the mean free paths of carriers are larger than the system size so that the transport is effectively ballistic. The impacts of impurities, disorders, and many-body interactions on the $y$-directional transverse transport are still open questions and the effect of  the ballistic-diffusion crossover on the AAR-induced Hall-like effects is unclear; these topics should deserve further detailed investigations. 
Also, it would be interesting to explore the possible quantization effect for transverse transport by constraining the $y$-direction within a finite small width. 
The effect of the possible {\it specular} AR~\cite{Beenakker,DYXing} on the Hall-like effects deserves attention in a future study.

\begin{acknowledgments}
{The work was supported by the National Nuclear Security Administration of the U.S. DOE at LANL under Contract No. DE-AC52-06NA25396, and through the LDRD Program of LANL. This work was supported, in part, by the Center for Integrated Nanotechnologies, a U.S. DOE user facility.}
\end{acknowledgments}


\begin{thebibliography}{01}
\bibitem{Andreev} A. F. Andreev, Zh. Eksp. Teor. Fiz. {\bf 46}, 1823 (1964) [Sov. Phys. JETP {\bf19}, 1228 (1964)].

\bibitem{ZBCP} C.-R. Hu, Phys. Rev. Lett. {\bf 72}, 1526 (1994).

\bibitem{BTK2} Y. Tanaka and S. Kashiwaya, Phys. Rev. Lett. {\bf 74}, 3451 (1995).

\bibitem{CAR1} J. M. Byers and M.E. Flatte, Phys. Rev. Lett. {\bf 74}, 306 (1995).
\bibitem{CAR2} G. Deutscher and D. Feinberg, Appl. Phys. Lett. {\bf 76}, 487 (2000).

\bibitem{Beenakker} C. W. J. Beenakker, Phys. Rev. Lett. {\bf 97}, 067007 (2006).
\bibitem{DYXing} Q. Zhang, D. Fu, B. Wang, R. Zhang, and D. Y. Xing, Phys. Rev. Lett. {\bf 101}, 047005 (2008).

\bibitem{Fu} L. Fu and C. L. Kane, Phys. Rev. Lett. {\bf 100}, 096407 (2008).


\bibitem{like-mean} Because the transverse transport effects induced by the asymmetric Andreev reflection do not have the same mechanism as the conventional Hall effects, which require time-reversal symmetry breaking by magnetic fields, throughout this work, we call Andreev-reflection-induced effects ``Hall-like'' effects.

\bibitem{Hasan2010RMP} M. Z. Hasan and C. L. Kane, Rev. Mod. Phys. {\bf82}, 3045 (2010).
\bibitem{Qi2011RMP} X.-L. Qi and S.-C. Zhang, Rev. Mod. Phys. {\bf83}, 1057 (2011).

\bibitem{savelev2012PRL} S. E. Savel'ev, W. H\"ausler, and P. H\"anggi, Phy. Rev. Lett. {\bf 109}, 226602 (2012).
\bibitem{HallTE} C. Zhou, S. Birner, Y. Tang, K. Heinselman, and M. Grayson, Phys. Rev. Lett. {\bf 110}, 227701 (2013). D. P. Monroe, Physics {\bf 6}, 63 (2013).

\bibitem{BdGbook} P. G. de Gennes, {\it Superconductivity of Metals and Alloys} (New York, Benjamin, 1989).
\bibitem{Tinkhambook} M. Tinkham, {\it Introduction to Superconductivity} (New York, Dover, 2004).


\bibitem{BTK00} J. Demes and A. Griffin, Can. J. Phys. {\bf 49}, 285 (1970).
\bibitem{BTK0} W. N. Mathews Jr., Phys. Stat. Sol. B {\bf90}, 327 (1978).

\bibitem{BTK1} G. E. Blonder, M. Tinkham, and T. M. Klapwijk, Phys. Rev. B {\bf 25}, 4515 (1982).

\bibitem{imperfectness} Conventionally, the imperfect interface effect with both barrier scattering and roughness scattering is traditionally renormalized into this imperfectness parameter $Z$; see, for example, Y.-H. Liao,  M. Yang, C. Ma, and Y.-B. Cao, Low Temp. Phys. {\bf38}, 368 (2012), or H. E. Camblong and  P. M. Levy, Phy. Rev. Lett. {\bf 69}, 2835 (1992).

\bibitem{Bhattacharjee2006PRL} S. Bhattacharjee and K. Sengupta, Phys. Rev. Lett. {\bf97}, 217001 (2006).

\bibitem{Ren2013PRB} J. Ren and J.-X. Zhu, Phys. Rev. B {\bf 87}, 165121 (2013).

\bibitem{Klein1} O. Klein, Z. Phys. {\bf 53}, 157 (1929)
\bibitem{Klein2} M. I. Katsnelson, K. S. Novoselov, and A. K. Geim, Nature Phys. {\bf 2}, 620 (2006).

\bibitem{heatBTK1} R. A. Riedel and P. F. Bagwell, Phys. Rev. B {\bf48}, 15198 (1993).
\bibitem{heatBTK2} A. Bardas and D. Averin, Phys. Rev. B {\bf 52}, 12873 (1995).

\bibitem{absent} It is worth noting that in the linear response regime, Nernst-Ettingshausen effects are conjugate to each other. A finite (no) Ettingshausen effect  will symmetrically indicate the same (no) Nernst effect, which is constrained by the so-called Onsager reciprocal relation. Our observation here of a finite Ettingshausen effect but no Nernst effect does not conflict with the Onsager relation, because the differential Hall conductance discussed here is a  nonlinear response quantity. When the bias $eV=E$ tends to 0, i.e. in the linear response regime, it is clear that the Ettingshausen effect described in Fig.~\ref{fig:nhall}(c) vanishes as well. If the electron-hole symmetry is broken in the present system, a finite AAR-induced Nernst-like effect will then be present, after the nonzero longitudinal thermal-induced electric current~\cite{Ren2013PRB}.

\bibitem{exp} See, for example, J. Y. T. Wei, N.-C. Yeh, D. F. Garrigus, and M. Strasik, Phys. Rev. Lett. {\bf 81}, 2542 (1998); S. Sinha and K.-W. Ng, Phys. Rev. Lett. {\bf80}, 1296 (1998).

\bibitem{JXZhu:1996} J.-X. Zhu, Z. D. Wang, and H. X. Tang
Phys. Rev. B {\bf 54}, 7354 (1996).

\bibitem{HXTang:1996} H. X. Tang, Z. D. Wang, and J.-X. Zhu
Phys. Rev. B {\bf 54}, 12509 (1996).

\bibitem{nano_exp} P. Rickhaus, M. Weiss, L. Marot, and C. Sch\"onenberger, Nano Lett. {\bf12}, 1942 (2012).


\end{thebibliography}
\end{document}